\documentclass[11pt] {article}
\usepackage{latexsym,amsfonts}
\begin{document}
\setcounter{page}{1}
\def\theequation{\arabic{section}.\arabic{equation}}
\def\theequation{\thesection.\arabic{equation}}
\setcounter{section}{0}

\title{On the dual--vector field condensation \\ in the dual Monopole
Nambu--Jona--Lasinio model with dual Dirac strings}

\author{M. Faber\thanks{E--mail: faber@kph.tuwien.ac.at, Tel.:
+43--1--58801--14261, Fax: +43--1--58801--14299} ,
A. N. Ivanov\thanks{E--mail: ivanov@kph.tuwien.ac.at, Tel.:
+43--1--58801--14261, Fax: +43--1--58801--14299}~$^{\S}$ ,
A. M\"uller\thanks{E--mail: mueller@kph.tuwien.ac.at, Tel.:
+43--1--58801--14265, Fax: +43--1--58801--14299} ,
N. I. Troitskaya\thanks{Permanent Address: State Technical University,
Department of Nuclear Physics, 195251 St. Petersburg, Russian
Federation}}

\date{\today}

\maketitle

\begin{center}
{\it Institut f\"ur Kernphysik, Technische Universit\"at Wien, \\
Wiedner Hauptstr. 8-10, A-1040 Vienna, Austria}
\end{center}

\vskip1.0truecm
\begin{center}
\begin{abstract}
The condensation of a dual--vector field is investigated in the dual
Monopole Nambu--Jona--Lasinio model with dual Dirac strings. The
condensate of a dual--vector field is calculated as a functional of a
shape of a dual Dirac string. The obtained result is compared with the
gluon condensate calculated in a QCD sum rules approach and on
lattice.
\end{abstract}
\end{center}

\newpage

\section{Introduction}
\setcounter{equation}{0}

\noindent{\bf Effective Lagrangian of the dual magnetic Monopole
Nambu--Jona--Lasinio model}. In our recent publications [1--4] we have
shown that the dual Monopole Nambu--Jona--Lasinio model (MNJL) with
dual Dirac strings is a perfect continuum space--time analogy of
Compact Quantum Electrodynamics (CQED) [5].  As has been shown in
Ref.[5] CQED possesses the same non--perturbative phenomena as
low--energy QCD. The MNJL model is based on a Lagrangian, invariant
under magnetic $U(1)$ symmetry, with massless magnetic monopoles
self--coupled through a local four--monopole interaction [1,2]:
\begin{eqnarray}\label{label1.1}
\hspace{-0.3in}{\cal L}(x) = \bar{\chi}(x) i \gamma^{\mu} \partial_{\mu}
 \chi(x) + G [\bar{\chi}(x) \chi(x)]^2 - G_1
[\bar{\chi}(x)\gamma_{\mu}\chi(x)][\bar{\chi}(x)\gamma^{\mu}\chi(x)],
\end{eqnarray}
where $\chi(x)$ is a massless magnetic monopole field, $G$ and $G_1$
are positive phenomenological constants. Below we will show that we
have to choose $G_1=G/4$ for the self consistency of the theory in the
one loop approximation [3,4].
 
The magnetic monopole condensation accompanies the creation of
$\bar{\chi} \chi$ collective excitations with the quantum numbers of
a scalar Higgs meson field $\rho$ and a dual--vector field $C_{\mu}$.

For the derivation of an effective Lagrangian the $\rho$ and $C_{\mu}$
fields are introduced as cyclic variables.
\begin{eqnarray}\label{label1.2}
& & {\cal L}(x)=\bar{\chi}(x)i\gamma^{\mu }\partial_{\mu}
\chi(x) - {\cal V}(x),
\end{eqnarray}
where ${\cal V}(x)$ is defined
\begin{eqnarray}\label{label1.3}
-{\cal V}(x) = \bar{\chi}(x)( -g\,\gamma^{\mu }C_{\mu}(x) -
 \kappa\,\rho (x))\,\chi (x) - \frac{\kappa^{2}}{4G}\rho^{2}(x) +
 \frac{g^{2}}{4G_{1}}C_{\mu }(x)C^{\mu}(x).
\end{eqnarray}
Now we can show that the vacuum expectation value (v.e.v) of the
$\rho$ field does not vanish.  For this aim we have to derive the
equation of motion of the $\rho$ field by varying the Lagrangian
Eq.(\ref{label1.1}) with respect to the $\rho$ field:
\begin{eqnarray}\label{label1.4}
\frac{\partial {\cal L}(x)}{\partial \rho (x)} = -\,\kappa\,
\bar{\chi}(x)\chi(x) - \frac{\kappa^2}{4G}\,\rho (x) = 0. 
\end{eqnarray}
This leads to
\begin{eqnarray}\label{label1.5}
\rho (x) = - \frac{2G}{\kappa}\,\bar{\chi}(x)\chi (x). 
\end{eqnarray}
Taking the v.e.v. of both sides of Eq.(\ref{label1.5}) we get
\begin{eqnarray}\label{label1.6}
\langle\rho (x)\rangle = -\,\frac{2G}{\kappa}\,\langle\bar{\chi}(x)\chi (x)\rangle =
-\,\frac{2G}{\kappa} <\,\!\bar{\chi}(0)\chi (0)\rangle ,
\end{eqnarray}
where $\langle\bar{\chi}(0) \chi(0) \rangle$ is the magnetic monopole
condensate. Thus, the non--zero value of the v.e.v. of the $ \rho $
field is related to the monopole condensation. In order to deal with a
physical scalar field, the $\sigma$--field, we should follow the
standard procedure and subtract $< \! \!\rho(x) \! \!>$. This gives
$\sigma(x) = \rho(x)\,- < \! \!\rho(x) \!  \!> $. The v.e.v. of the
$\rho$--field can be expressed in terms of the mass of the magnetic
monopole $M$ in the superconducting phase [1--4], $< \!\!\rho(x) \! \!>
= M/\kappa$, where $M$ is proportional to $< \!  \!\bar{\chi}(0)
\chi(0) \! \!>$ [1--4]
\begin{eqnarray}\label{label1.7}
M = -\,2\,G\,\langle\bar{\chi}(0) \chi(0)\rangle.
\end{eqnarray}
This is the gap--equation testifying the appearance of massive
 magnetic monopoles in the superconducting phase, where $<\!
 \!\bar{\chi}(0) \chi(0)\! \!> \not= 0$. As has been shown in
 Refs.[1,2] it also leads to the suppression of direct transitions
 between the physical scalar field $\sigma$ and the non--perturbative
 vacuum.

In terms of the $\sigma$--field the Lagrangian Eq.(\ref{label1.2}) reads
\begin{eqnarray}\label{label1.8}
{\cal L}(x) = \bar{\chi}(x)( i\gamma^{\mu}\partial _{\mu } - M )\,
\chi (x) - {\tilde{\cal V}}(x),
\end{eqnarray}
where now ${\tilde{\cal V}}(x)$ reads
\begin{eqnarray}\label{label1.9}
- {\tilde{\cal V}}(x)=\bar{\chi}(x)( -\,g\,\gamma^{\mu }C_{\mu}(x) -
 \kappa\, \sigma (x)) \chi (x) - \frac{\kappa^2}{4G}\rho^{2}(x) +
 \frac{g^2}{4G_{1}}C_{\mu }(x)C^{\mu }(x) .
\end{eqnarray}
Integrating out the magnetic monopole fields we arrive at the
effective Lagrangian
\begin{eqnarray}\label{label1.10}
{\cal L}_{\rm eff}(x) = {\tilde{\cal L}}_{\rm eff} -
\frac{\kappa^{2}}{4G}\rho^2(x) + \frac{g^2}{4G_1}\,C_{\mu }(x)C^{\mu}(x)
\end{eqnarray}
with ${\tilde{\cal L}}(x)_{\rm eff}$ defined as
\begin{eqnarray}\label{label1.11}
{\tilde{\cal L}}_{\rm eff}(x)\,=\,-\,i\,\Bigg<\,x\Bigg|{\ell n}\frac{{\rm
Det}(i\,\hat{\partial}\,-\,M\,+\,\Phi)}{{\rm
Det}(i\,\hat{\partial}\,-\,M)}\Bigg|x\,\Bigg>\,.
\end{eqnarray}
Here we have denoted $\Phi = - g \gamma^{\mu} C_{\mu} - \kappa \sigma$, and
$\sigma = \rho\,-\, M/\kappa$.

The effective Lagrangian ${\tilde{\cal L}}_{\rm eff}(x)$ can be represented
by an infinite series
\begin{eqnarray}\label{label1.12}
{\tilde{\cal L}}_{\rm eff}(x) = \sum_{n = 1}^{\infty}\frac{i}{n}\,{\rm
tr}_{\,\rm L}\Bigg< x
\Bigg|\Bigg(\frac{1}{M\,-\,i\,\hat{\partial}}\,\Phi\Bigg)^n
\Bigg|x\,\Bigg>\,=\,\sum_{n = 1}^{\infty}{\tilde{\cal L}}^{(n)}_{\rm
eff}(x) \,.
\end{eqnarray}
The index ${\rm L}$ means the evaluation of the trace over the Lorentz
indices. The effective Lagrangian ${\tilde{\cal L}}^{(n)}_{\rm eff}(x)$
is given by
\begin{eqnarray}\label{label1.13}
\tilde{{\cal L}}^{(n)}_{\rm eff} (x)&=&\int\,\prod_{\ell
=\,1}^{n\,-\,1}\frac{d^4 x_{\ell} d^4 k_{\ell}}{(2 \pi)^4}\,
e^{\displaystyle\,-\,i\,k_1\cdot x_1\,-\ldots -i\,k_n\cdot
x\,}\,(-\,\frac{1}{n}\,\frac{1}{16 \pi^2})\,\int \frac{d^4
k}{\pi^2 i}\nonumber\\ &&\times{\rm tr}_{\,\rm
L}\,\Bigg\{\,\frac{1}{M\,-\,\hat{k}}\,\Phi\,(x_1)\,\frac{1}{M\,-\,
\hat{k}\,-\,\hat{k}_1}\,\Phi
(x_2)\,\ldots \nonumber\\ && \times \ldots
\Phi\,(x_{\,n\,-\,1})\,\frac{1}{M\,-\,\hat{k}\,-\,\hat{k}_1\,-\ldots
-\,\hat{k}_{n-1}}\,\Phi (x)\,\Bigg\}\,.
\end{eqnarray}
at $\,k_1\,+\,k_2\,+\ldots +\,k_n\,=\,0\,$. The r.h.s. of
Eq.(\ref{label1.13}) describes the one--massive--monopole loop diagram with
$n$--vertices. The monopole--loop diagrams with two vertices
$(n\,=\,2)$ determine the kinetic term of the $\sigma$--field and give
the contribution to the kinetic term of the $C_{\mu}$--field, while the
diagrams with ($n\ge\,3$) describe the vertices of interactions of the
$\sigma$ and $C_{\mu}$ fields. In accordance with the prescription
given in [1,2] the effective Lagrangian $\tilde{{\cal L}}_{\rm eff} (x)$
should be  defined by the set of divergent one-massive--monopole--loop
diagrams with $n\,=\,1,2,3$ and $4$ vertices. The evaluation of these
diagrams gives
\begin{eqnarray}\label{label1.14}
{\cal L}_{\rm
eff}\,(x)&=&\frac{1}{2}\,\frac{\kappa^2}{8\,\pi^2}\,J_2(M)\,\partial_{\mu}\,
\sigma(x)\,\partial^{\mu}\,\sigma(x)\,-\,M\,\Bigg[\frac{\kappa}{2\,G}\,-\,
\frac{\kappa}{4\,\pi^2}\,J_1(M)\Bigg]\,\sigma(x)\,+ \nonumber\\
&&+\,\frac{1}{2}\Bigg[-\,\frac{\kappa^2}{2\,G}\,+\,\frac{\kappa^2}{4\,\pi^2}
J_1(M)\,-\,4\,M^2\,\frac{\kappa^2}{8\,\pi^2}\,J_2(M)\Bigg]\,\sigma^2(x)\,
\nonumber\\
&&-\,2\,M\,\kappa\,\frac{\kappa^2}{8\,\pi^2}\,J_2(M)\,\sigma^3(x)\,-\,
\frac{1}{2}\,\kappa^2\,\frac{\kappa^2}{8\,\pi^2}\,J_2(M)\,\sigma^4(x)\,
\nonumber\\ &&-\frac{g^2}{48\,\pi^2}\,J_2(M)\, C_{\mu\nu}(x)\,
C(x)^{\mu\nu}\,\nonumber\\ &&+\Bigg[\frac{g^2}{4
G_1}-\frac{g^2}{16\,\pi^2}\,[J_1(M)\,+\,M^2\,J_2(M)]\Bigg]\,
C_{\mu}(x)\,C^{\mu}(x)\,,
\end{eqnarray}
where we have defined $C^{\mu\nu}(x) = \partial^{\mu} C^{\nu}(x) -
\partial^{\nu} C^{\mu}(x)$. Then, $J_1(M)$ and $J_2(M)$ are the
following quadratically and logarithmically divergent integrals
\begin{eqnarray}\label{label1.15}
J_1(M)\,=\,\int\,\frac{d^4 k}{\pi^2 i}\,\frac{1}{(M^2\,-\,k^2)} = 
\Lambda^2 - M^2{\ell n}\Bigg(1 + \frac{\Lambda^2}{M^2}\Bigg)\,-\,
\frac{\Lambda^2}{M^2\,+\,\Lambda^2},\nonumber\\
J_2(M)\,=\,\int\,\frac{d^4 k}{\pi^2 i}\,\frac{1}{(M^2\,-\,k^2)^2} = 
{\ell n}\Bigg(1 + \frac{\Lambda^2}{M^2}\Bigg)\,-\,
\frac{\Lambda^2}{M^2\,+\,\Lambda^2},
\end{eqnarray}
In order to get correct factors of the $\sigma$ and $C_{\mu}$ field kinetic
terms we have to set [1,2]
\begin{eqnarray}\label{label1.16}
\frac{g^2}{12\,\pi^2}\,J_2(M)\,=\,1 \quad,\quad
\frac{\kappa^2}{8\,\pi^2}\,J_2(M)\,=\,1\,.
\end{eqnarray}
So, the coupling constants are connected by the relation
$\kappa^2 = 2g^2/3$ [1,2].

The effective Lagrangian Eq.(\ref{label1.14}) contains a term linear
in the $\sigma$--field. This part of the effective Lagrangian leads to
direct transitions $\sigma \to $ vacuum. In the case of a physical
$\sigma$--field such transitions should be suppressed. In order to
suppress these transitions we have to impose the constraint [1--4]
\begin{eqnarray}\label{label1.17}
\frac{1}{G} - \frac{1}{2\pi^2}\,J_1(M) = 0,
\end{eqnarray}
where $J_1(M)$ can be  connected with the monopole condensate [1--4]
\begin{eqnarray}\label{label1.18}
<\bar{\chi}(0) \chi(0)> = -\,\frac{1}{4\pi^2}MJ_1(M). 
\end{eqnarray}
Inserting Eq.(\ref{label1.18}) into Eq.(\ref{label1.17}) we arrive at
the gap--equation (\ref{label1.7})

The coefficient in front of the last term in Eq.(\ref{label1.14})
defines the mass of the $C_{\mu}$ field:
\begin{eqnarray}\label{label1.19}
M^2_C = \frac{g^2}{2G_1} - \frac{g^2}{8\,\pi^2}\,[J_1(M)\,+\,M^2\,J_2(M)].
\end{eqnarray}
Piling up the gap--equation (\ref{label1.7}) and the constraint
(\ref{label1.16}) we recast the effective Lagrangian (\ref{label1.14})
into the form
\begin{eqnarray}\label{label1.20}
{\cal L}_{\rm eff}(x)&=&-\frac{1}{4}\,C_{\mu\nu}(x)\,C^{\mu\nu}(x) \, +
\,\frac{1}{2}\,M^2_C\,C_{\mu}(x)\,C^{\mu}(x)\,+\nonumber\\
&&+\frac{1}{2} \partial_{\mu}\,\sigma(x)\,\partial^{\mu} \sigma(x) -
\frac{1}{2} M^2_{\sigma} \sigma^2(x) \Bigg[1 + \kappa
\frac{\sigma(x)}{M_{\sigma}}\Bigg]^2\,,\nonumber\\
&=&-\frac{1}{4}\,dC_{\mu\nu}(x)\, dC^{\mu\nu}(x) \, +
\,\frac{1}{2}\,M^2_C\,C_{\mu}(x)\,C^{\mu}(x)\,+\nonumber\\
&&+\frac{1}{2} \partial_{\mu}\,\sigma(x)\,\partial^{\mu} \sigma(x) -
\frac{1}{2} M^2_{\sigma} \sigma^2(x) + {\cal L}_{\rm int}[\sigma(x)],
\end{eqnarray}
where $M_{\sigma}\,=\,2\,M$ is the mass of the $\sigma$--field and
${\cal L}_{\rm int}[\sigma(x)] $ describes the self--interactions of
the $\sigma$--field
\begin{eqnarray}\label{label1.21}
{\cal L}_{\rm int}[\sigma(x)] = - \kappa\,M_{\sigma}\,\sigma^3(x) -
\frac{1}{2}\kappa^2\,\sigma^4(x).
\end{eqnarray}
As has been shown in Ref.[3,4] the gap--equation, in turn, can be
derived in the one--monopole loop approximation by using only the
Lagrangian Eq.(\ref{label1.1}). When comparing these two
gap--equations we get the relation $G_1 = G/4$ that reduces the number
of input parameters [3,4].

\noindent{\bf Wave function of the non--perturbative vacuum}. The MNJL
model as well as the NJL model [6] and the BCS theory of superconductivity [7]
possesses a non--trivial non--perturbative vacuum with a wave function
[2]
\begin{eqnarray}\label{label1.22}
|0\rangle^{(M)} =
 \prod_{\vec{p},\lambda=\pm1}\Bigg[\sqrt{\frac{1+\beta_{\vec{p}}}{2}}
 + \lambda\,\sqrt{\frac{1-\beta_{\vec{p}}}{2}}\,a^{(0)\dagger}(\vec{p},\lambda)\,b^{(0)\dagger}(- \vec{p},\lambda)\Bigg]\,|0\rangle,
\end{eqnarray}
where $\beta_{\vec{p}} = \vec{p}/E_{\vec{p}} =
\vec{p}/\sqrt{\vec{p}^{\,2} + M^2}$ is the velocity of a massive
monopole with the mass $M$, and $a^{(0)\dagger}(\vec{p},\lambda)\,$
(or $b^{(0)\dagger}(- \vec{p},\lambda)$) denotes the creation operator
of a massless monopole ( or anti--monopole) with a momentum $\vec{p}$
and helicity $\lambda$; $|0\rangle = |0\rangle^{(0)}$ is the wave function
of the perturbative vacuum of the non--condensed phase. The wave
function $|0\rangle^{(M)}$ of the non--perturbative vacuum is distinctly
invariant under magnetic $U(1)$ symmetry. The former implies that in
the condensed phase the magnetic $U(1)$ symmetry is not broken
[2]. This is the main peculiarity of the MNJL model with respect to
the dual Higgs model with dual Dirac strings [8--10], where the magnetic
$U(1)$ symmetry is broken spontaneously in the superconducting phase.

The wave function of the non--perturbative vacuum $|0\rangle^{(M)}$ is
invariant under party transformations ${\cal P}$, ${\cal P}
|0\rangle^{(M)}=|0\rangle^{(M)}$.

In order to show that the magnetic monopole condensate
$\langle\bar{\chi}(0)\chi(0)\rangle$ has a distinct meaning of the
order parameter we, following the BCS theory of superconductivity [7],
can introduce two operators possessing different properties under
parity transformations
\begin{eqnarray}\label{label1.23}
{\cal O}_+ &=& 2\sum_{\vec{p}}\sum_{\lambda = \pm
 1}\lambda\,b^{(0)\dagger}(-
 \vec{p},\lambda)\,a^{(0)\dagger}(\vec{p},\lambda)\,,\,{\cal P} {\cal
 O}_+ {\cal P}^{\dagger} = +\,{\cal O}_+,\nonumber\\ {\cal O}_- &=&
 2\sum_{\vec{p}}\sum_{\lambda = \pm 1}~~b^{(0)\dagger}(-
 \vec{p},\lambda)\,a^{(0)\dagger}(\vec{p},\lambda)\,,\,{\cal P} {\cal
 O}_- {\cal P}^{\dagger} = -\,{\cal O}_-.
\end{eqnarray}
As has been shown in Ref.[2] the v.e.v of the ${\cal O}_+$ operator
per unit volume coincides with the magnetic monopole condensate
$\langle{\cal O}_+\rangle = \langle\bar{\chi}(0)\chi(0)\rangle$:
\begin{eqnarray}\label{label1.24}
\hspace{-0.5in}&&\langle{\cal O}_+\rangle = \frac{1}{V}
 {^{(M)}\langle 0|{\cal O}_+|0\rangle^{(M)}} =
 -\,\frac{1}{V}\sum_{\vec{p}}\sum_{\lambda = \pm 1}\lambda^2 \sqrt{1 -
 \beta^2_{\vec{p}}} = - 4 M \sum_{\vec{p}}\frac{1}{2E_{\vec{p}}V} =
 \nonumber\\ 
\hspace{-0.5in}&&= - 4 M \int \frac{d^3 p}{2E_{\vec{p}}} = - 4 M \int
\frac{d^4 p}{(2\pi)^4 i}\,\frac{1}{M^2 - p^2 - i\,0} =
-\,\frac{M}{4\pi^2}\,J_1(M) = \langle\bar{\chi}(0)\chi(0)\rangle,
\end{eqnarray}
where $V$ is a normalization volume. In turn, one can show that
$\langle{\cal O}_-\rangle = 0$. This confirms the parity conservation
in the MNJL model.

\noindent{\bf Dual Dirac strings}. Dual Dirac strings are included in
the MNJL model in the form of a dual electric field strength ${\cal
E}^{\mu\nu}(x)$ defined by [1--4]
\begin{eqnarray}\label{label1.25}
\hspace{-0.2in}&&{\cal E}^{\mu\nu}(x) = Q \int\!\!\!\int  d\tau d \sigma
\Bigg(\frac{\partial X^{\mu}}{\partial \tau}
\frac{\partial X^{\nu}}{\partial \sigma} -
\frac{\partial X^{\nu}}{\partial \tau}
\frac{\partial X^{\mu}}{\partial \sigma}\Bigg)\delta^{(4)}(x - X),
\end{eqnarray}
where $X^{\mu} = X^{\mu} (\tau,\sigma)$ represents the position of a point
on the world sheet swept by the string. The sheet is parameterized by
internal coordinates  ${-\infty} < \tau < {\infty}$ and $0 \le \sigma \le
\pi$, so that $X^{\mu} (\tau, 0) = X^{\mu}_{ - Q}(\tau)$ and
$X^{\mu}(\tau,\pi) = X^{\mu}_Q(\tau)$ represent the world lines of an
anti--quark and a quark [1--4,8--10].

The effective Lagrangian of the dual Higgs field $\sigma(x)$ and
vector $C^{\mu}(x)$ fields is then defined
\begin{eqnarray}\label{label1.26}
{\cal L}_{\rm eff}(x)&=&\frac{1}{4}\,F_{\mu\nu}(x)\, F^{\mu\nu}(x) \, +
\,\frac{1}{2}\,M^2_C\,C_{\mu}(x)\,C^{\mu}(x)\,+\nonumber\\
&&+\frac{1}{2} \partial_{\mu}\,\sigma(x)\,\partial^{\mu} \sigma(x) -
\frac{1}{2} M^2_{\sigma} \sigma^2(x) \Bigg[1 + \kappa
\frac{\sigma(x)}{M_{\sigma}}\Bigg]^2\,,\nonumber\\
&=&\frac{1}{4}\,F_{\mu\nu}(x)\, F^{\mu\nu}(x) \, +
\,\frac{1}{2}\,M^2_C\,C_{\mu}(x)\,C^{\mu}(x)\,+\nonumber\\
&&+\frac{1}{2} \partial_{\mu}\,\sigma(x)\,\partial^{\mu} \sigma(x) -
\frac{1}{2} M^2_{\sigma} \sigma^2(x) + {\cal L}_{\rm int}[\sigma(x)],
\end{eqnarray}
where the field strength $F^{\mu\nu}(x)$ reads [1--4,8--10]:
$F^{\mu\nu}(x) = {\cal E}^{\mu\nu}(x) - {^*C^{\mu\nu}(x)}$ and
${^*C^{\mu\nu}(x)}$ is the dual version of $C^{\mu\nu}(x)$,
${^*C^{\mu\nu}(x)} = \frac{1}{2} \varepsilon^{\mu\nu\alpha\beta}
C_{\alpha\beta}(x)\,(\varepsilon^{0123} = 1)$.

\noindent{\bf Quarks and anti--quarks}. In the MNJL model a quark and
an anti--quarks are point--like particles with masses $m_q =
m_{\bar{q}} = m$, electric charges $Q_q = - Q_{\bar{q}} = Q$, and
trajectories $X^{\nu}_q(\tau)$ and $X^{\nu}_{\bar{q}}(\tau)$,
respectively, attached to the ends of a dual Dirac string. They are
described by the Lagrangian
\begin{eqnarray}\label{label1.27}
\hspace{-0.3in}{\cal L}_{\rm free~quark}(x) = - \sum_{i=q,\bar{q}} m_i
 \int d{\tau} \Bigg(\frac{d X^{\mu}_i(\tau)}{d\tau} 
\frac{dX^{\nu}_i(\tau)}{d\tau} g_{\mu\nu} \Bigg)^{1/2}
\delta^{(4)} (x - X_i(\tau)).
\end{eqnarray}
Substituting Eq.(\ref{label1.27}) in Eq.(\ref{label1.26}) we arrive at
the total effective Lagrangian of the MNJL model with dual Dirac
strings, quarks and anti--quarks
\begin{eqnarray}\label{label1.28}
{\cal L}_{\rm eff}(x)&=&\frac{1}{4}\,F_{\mu\nu}(x)\, F^{\mu\nu}(x) \,
+ \,\frac{1}{2}\,M^2_C\,C_{\mu}(x)\,C^{\mu}(x)\,+\nonumber\\
&&+\frac{1}{2} \partial_{\mu}\,\sigma(x)\,\partial^{\mu} \sigma(x) -
\frac{1}{2} M^2_{\sigma} \sigma^2(x) \Bigg[1 + \kappa
\frac{\sigma(x)}{M_{\sigma}}\Bigg]^2\,,\nonumber\\
&=&\frac{1}{4}\,F_{\mu\nu}(x)\,F^{\mu\nu}(x) \, +
\,\frac{1}{2}\,M^2_C\,C_{\mu}(x)\,C^{\mu}(x)\,+\nonumber\\
&&+\frac{1}{2} \partial_{\mu}\,\sigma(x)\,\partial^{\mu} \sigma(x) -
\frac{1}{2} M^2_{\sigma} \sigma^2(x) + {\cal L}_{\rm
int}[\sigma(x)]\nonumber\\ 
&&+ {\cal L}_{\rm free~quark}(x).
\end{eqnarray}
This effective Lagrangian should be used for the evaluation of
different observables defined in terms of vacuum expectation values related to the magnetic monopole Green functions [1--4].

\noindent{\bf Magnetic monopole Green functions}. The $n$--point
magnetic monopole Green function can be defined as the vacuum
expectation value of the time ordered product of the massless magnetic
monopole densities Refs.[1--4]:
\begin{eqnarray}\label{label1.29}
G\,(x_{1},\ldots, x_{n}) = \langle 0|{\rm T}(\bar{\chi} (x_{1}) \Gamma_{1}
 \chi (x_{1}) \ldots \bar{\chi} (x_{n})
\,\Gamma_{n} \chi\,(x_{n}))|0 \rangle_{\rm conn.}\,,
\end{eqnarray}
where $\Gamma_i (i = 1,\ldots,n)$ are the Dirac matrices. As has been
shown in Ref.[11] the vacuum expectation value Eq.(\ref{label1.29})
can be represented in terms of the vacuum expectation values of the
densities of the massive magnetic monopole fields $\chi_M(x)$ coupled
to the fields of the collective excitations $\sigma$ and $C_{\mu}$
\begin{eqnarray}\label{label1.30}
\hspace{-0.5in}&&G(x_{1},\ldots,x_{n})=
\langle 0|{\rm T}(\bar{\chi}(x_{1}) \Gamma_1 \chi(x_{1}) \ldots \bar{\chi}(x_{n})
\Gamma_n \chi(x_{n}))|0\rangle_{\rm conn.} = \nonumber\\
\hspace{-0.3in}&&= ^{(M)}\langle 0|{\rm T} \Big( \bar{\chi}_M(x_{1}) \Gamma_1
\chi_M(x_{1}) \ldots \bar{\chi}_M(x_{n}) \Gamma_{n} \chi_M(x_{n})
\nonumber\\
\hspace{-0.5in}&&\times\,\exp i\int d^4x \{-g\bar{\chi}_M(x)
\gamma^{\nu} \chi_M(x) C_{\nu}(x)- \kappa \bar{\chi}_M(x) \chi_M(x)
\sigma(x) + {\cal L}_{\rm int}[\sigma(x)]\}\Big)|0\rangle^{(M)}_{\rm
conn.}.\nonumber\\ &&
\end{eqnarray}
Here $|0\rangle^{(M)}$ is the wave function of the non--perturbative vacuum of
the MNJL model in the condensed phase and $|0\rangle$ the wave function of the
perturbative vacuum of the non--condensed phase. 

The self--interactions ${\cal L}_{\rm int}[\sigma(x)]$ provide
$\sigma$--field loop contributions and can be dropped out in the tree
$\sigma$--field approximation Refs. [1--4]. The tree $\sigma$--field
approximation can be justified keeping massive magnetic monopoles very
heavy, i.e. $M \gg M_C$. This corresponds to the London limit
$M_{\sigma} = 2\,M \gg M_C$ in the dual Higgs model with dual Dirac
strings [8--10]. The inequality $M_{\sigma} \gg M_C$ means also that
in the MNJL model we deal with {\it Dual Superconductivity of type
{\rm I$\!$I}} [12]. In the tree $\sigma$--field approximation the
r.h.s. of Eq.(\ref{label1.30}) can be recast into the form
\begin{eqnarray}\label{label1.31}
\hspace{-0.3in}&&G(x_{1},\ldots,x_{n})=<{0}|{\rm T}(\bar{\chi}(x_{1})
 \Gamma_1 \chi(x_{1}) \ldots \bar{\chi}(x_{n}) \Gamma_n
 \chi(x_{n}))|{0}>_{\rm conn.} = \nonumber\\
\hspace{-0.3in}&&= ^{(M)}\!<{0}|{\rm T} \Big( \bar{\chi}_M(x_{1}) \Gamma_1
\chi_M(x_{1}) \ldots \bar{\chi}_M(x_{n}) \Gamma_{n} \chi_M(x_{n})
 \nonumber\\
\hspace{-0.3in}&&\exp i\int d^4x \Big\{-g\bar{\chi}_M(x)
 \gamma^{\nu} \chi_M(x) C_{\nu}(x) - \kappa \bar{\chi}_M(x) \chi_M(x)
\sigma(x)\}\Big)|{0}>^{(M)}_{\rm conn.}.
\end{eqnarray}
For the subsequent investigation it is convenient to represent the r.h.s.
of Eq.(\ref{label1.31}) in terms of the generating functional of the
monopole Green functions [1--4]
\begin{eqnarray}\label{label1.32}
\hspace{-0.3in}&&G(x_{1},\ldots,x_{n}) =
\prod^{n}_{i=1}\frac{\delta}{\delta \eta(x_i)}\Gamma_i
 \frac{\delta}{\delta \bar{\eta}(x_i)}Z[\eta,\bar{\eta}]
 \Bigg|_{\eta =\bar{\eta}=0},
\end{eqnarray}
where $\bar{\eta}(\eta)$ are the external sources of the massive monopole
(antimonopole) fields, and $Z[\eta,\bar{\eta}]$ is the generating
functional of the monopole Green functions defined by
\begin{eqnarray}\label{label1.33}
\hspace{-0.3in}&&Z[\eta,\bar{\eta}]=\frac{1}{Z}\int {\cal D}\chi_M {\cal
D}\bar{\chi}_M {\cal D}C_{\mu} {\cal D}\sigma\,\exp i\int d^4x\,\Big[
\frac{1}{4}\,F_{\mu\nu}(x)\,F^{\mu\nu}(x)\nonumber\\
\hspace{-0.3in}&&+ \frac{1}{2}\,M^2_C\,C_{\mu}(x)\,C^{\mu}(x)
+ \frac{1}{2}\,\partial_{\mu}\sigma(x)\,\partial^{\mu} \sigma(x)
  - \frac{1}{2}\,M^2_{\sigma}\,\sigma^2(x)\nonumber\\
\hspace{-0.3in}&&+ \bar{\chi}_M(x)(i\,\gamma^{\mu}\,\partial_{\mu} - M
 - g\,\gamma^{\mu}\,C_{\mu}(x) - \kappa\,\sigma(x))\,\chi_M(x)
  \nonumber\\
\hspace{-0.5in}&&+ \bar{\eta}(x)\,\chi_M(x) + \bar{\chi}_M(x)\,\eta(x)
+ {\cal L}_{\rm free~quark}(x)\Big].
\end{eqnarray}
The normalization factor $Z$ is defined by the condition $Z[0,0] = 1$.

The paper is organized as follows. In Sect.\,2 we discuss the gluon
condensate in non--perturbative QCD both in the QCD sum rules approach
and in lattice simulations. In Sect.\,3 we calculate the dual--vector
field condensate and compare the obtained result with the gluon
condensate of non--perturbative QCD. In the Conclusion we discuss the
obtained results.

\section{Gluon condensate of non--perturbative QCD}
\setcounter{equation}{0}

For the first time, the condensate of the gluon fields
\begin{eqnarray}\label{label2.1}
\Big\langle
\frac{g^2_s}{4\pi^2}\,G^a_{\mu\nu}(0)\,G^{a\mu\nu}(0)\Big\rangle
\not= 0,
\end{eqnarray}
where $g_s$ and $G^a_{\mu\nu}(0)\,(a = 1,\ldots,8)$ are the
quark--gluon coupling constant and the gluon field strength, has been
introduced  as a phenomenological parameter characterizing
quantitatively non--trivial properties of a non--perturbative vacuum
of QCD in the QCD sum rules approach [13]. The gluon condensate breaks
the dilatation invariance at a scale of energy transferred of order
$\Lambda_{\rm D} \sim 4\,{\rm GeV}$ [14]. This gives a signal to the
breaking of chiral symmetry which becomes broken spontaneously at a
scale of energy transferred of order $\Lambda_{\chi} \sim 1\,{\rm
GeV}$ [14]. As has been shown in Ref.[15] the contribution of the
gluon condensate to the condensate of light quarks $u$, $d$ and $s$
makes up 1/3 of the meanvalue of the quark condensate
$\langle\bar{u}u\rangle = \langle\bar{d}d\rangle =
\langle\bar{s}s\rangle = -(0.253\,{\rm GeV})^3$ [15].

The numerical value of the gluon condensate 
\begin{eqnarray}\label{label2.2}
\Big\langle
\frac{g^2_s}{4\pi^2}\,G^a_{\mu\nu}(0)\,G^{a\mu\nu}(0)\Big\rangle
 = (0.331\,{\rm GeV})^4
\end{eqnarray}
obtained in Ref.[13] by studying the charmonium channel has been
obviously underestimated [16--18]. The correct value of the gluon
condensate increased by a factor of 4 has been obtained in
Ref.[19]. The meanvalue of the gluon condensate is equal to [19]
\begin{eqnarray}\label{label2.3}
\Big\langle
\frac{g^2_s}{4\pi^2}\,G^a_{\mu\nu}(0)\,G^{a\mu\nu}(0)\Big\rangle =
(0.458\,{\rm GeV})^4.
\end{eqnarray}
In the dilute--instanton gas approximation the gluon condensate has
been defined as [13]
\begin{eqnarray}\label{label2.4}
\Big\langle
\frac{g^2_s}{4\pi^2}\,G^a_{\mu\nu}(0)\,G^{a\mu\nu}(0)\Big\rangle_{\rm inst. + anti-inst.} =
16\int\limits^{\textstyle \rho_c}_0\frac{d\rho}{\rho^5}\,d(\rho),
\end{eqnarray}
where $d(\rho)$ is the instanton density function. For the $SU(N)$
gauge group the instanton density function is defined by [20]
\begin{eqnarray}\label{label2.5}
d(\rho) =
\frac{C_1e^{\textstyle - C_2 N}}{(N-1)!(N-2)!}\,\Bigg[\frac{8\pi^2}{g^2_s(\rho)}\Bigg]^{2N}\,e^{\textstyle
- 8\pi^2/g^2_s(\rho)}.
\end{eqnarray}
The coefficients $C_1$ and $C_2$ are given by [20]
\begin{eqnarray}\label{label2.6}
C_1 &=& \frac{2\,e^{5/6}}{\pi^2} = 0.466,\nonumber\\ C_2 &=&
\frac{5}{3}\,{\ell n}\,2 - \frac{11}{36} + \frac{1}{3}\,({\ell
n}\,(2\pi) + \gamma) + \frac{2}{\pi^2}\sum^{\infty}_{n=0}\frac{{\ell
n}\,n}{n^2} = 1.679,
\end{eqnarray}
where $\gamma = 0.5277\ldots$ is Euler's constant.

In the absence of quark contributions and in the one--loop
approximation the running coupling constant $g(\rho)$ is determined
[20]
\begin{eqnarray}\label{label2.7}
g^2_s(\rho) = \frac{g^2_s(\rho_0)}{\displaystyle 1 +
\frac{11}{3}\,N\,\frac{g^2_s(\rho_0)}{8\pi^2}\,{\ell n}\frac{\rho}{\rho_0}},
\end{eqnarray}
where $\rho_0 = 1/\Lambda_{\rm U}$ and $\Lambda_{\rm U}$ is the
ultra--violet cut--off. Then, the integral over $\rho$ defining the
gluon condensate Eq.(\ref{label2.4}) is infrared divergent and the
parameter $\rho_c = 1/\Lambda_{\rm R}$ plays the role of the infrared
cut--off [13]. According to estimates of Ref.[13] the infrared
cut--off should of order $\Lambda_{\rm R}\sim 200\,{\rm MeV}$.

Thus, in the absence of quark contributions and in the one--loop
approximation the instanton density function $d(\rho)$ reads [20]
\begin{eqnarray}\label{label2.8}
d(\rho)= \frac{C_1\,e^{\textstyle - C_2
N}}{(N-1)!(N-2)!}\,(\rho\Lambda_{\rm U})^{11N/3}\Bigg[\frac{8\pi^2}{g^2_s(\Lambda_{\rm U})} -
\frac{11}{3}\,N\,{\ell
n}(\rho\Lambda_{\rm U})\Bigg]^{2N}\,e^{\textstyle
- 8\pi^2/g^2_s(\Lambda_{\rm U})}.
\end{eqnarray}
For QCD when $N=3$ we obtain
\begin{eqnarray}\label{label2.9}
d(\rho) = 1.513\,10^{-3}\,(\rho\Lambda_{\rm
U})^{11}\Bigg[\frac{8\pi^2}{g^2_s(\Lambda_{\rm U})} - 11\,{\ell
n}(\rho\Lambda_{\rm U})\Bigg]^6\,e^{\textstyle -
8\pi^2/g^2_s(\Lambda_{\rm U})}.
\end{eqnarray}
Substituting Eq.(\ref{label2.9}) in Eq.(\ref{label2.4}) one can see
that the integral over $\rho$ is concentrated near the infrared
cut--off $\rho_c=1/\Lambda_{\rm R}$. Therefore, the result of the
integration can be given by the expression [13,20]
\begin{eqnarray}\label{label2.10}
\Big\langle
\frac{g^2_s}{4\pi^2}\,G^a_{\mu\nu}(0)\,G^{a\mu\nu}(0)\Big\rangle_{\rm inst. + anti-inst.}
&\approx& 1.513\,10^{-3}\,\frac{16}{7}\,\Lambda^4_{\rm
R}\,\Bigg[\frac{8\pi^2}{g^2_s(\Lambda_{\rm R})} \Bigg]^6\,e^{\textstyle -
8\pi^2/g^2_s(\Lambda_{\rm R})} \nonumber\\ &\approx&
3.458\,10^{-3}\,\Lambda^4_{\rm R}\,\Bigg[\frac{8\pi^2}{g^2_s(\Lambda_{\rm R})}
\Bigg]^6\,e^{\textstyle - 8\pi^2/g^2_s(\Lambda_{\rm R})}.
\end{eqnarray}
According to lattice formulation of QCD the gluon condensate can be
represented in the following general form [21]
\begin{eqnarray}\label{label2.11}
\frac{1}{\Lambda^4_{\rm U}}\Big\langle
\frac{g^2_s}{4\pi^2}\,G^a_{\mu\nu}(0)\,G^{a\mu\nu}(0)\Big\rangle =
W_{40} + W_{04}\,\frac{\Lambda^4_{\rm QCD}}{\Lambda^4_{\rm U}} +
W_{22}\,\frac{\Lambda^2_{\rm QCD}}{\Lambda^2_{\rm U}} + \ldots,
\end{eqnarray}
where $W_{mn}$ are the numerical coefficients and $\Lambda_{\rm QCD}$ enters to the definition of the running coupling constant [19,22]
\begin{eqnarray}\label{label2.12}
\alpha_s(\rho) = \frac{g^2_s(\rho)}{4\pi} = 
\frac{4\pi}{b_0}\,\frac{1}{\displaystyle {\ell
n}\frac{1}{\rho \Lambda_{\rm QCD}}},
\end{eqnarray}
where $b_0 = (11/3)N$ [22]. The typical value of $\Lambda_{\rm QCD}$
is $\Lambda_{\rm QCD} = 100 \div 300\,{\rm MeV}$ [22].

The coefficient $W_{40}$ describes a perturbative contribution to the
gluon condensate, whilst the coefficients $W_{40}$ and $W_{22}$ are
fully non--perturbative. The presence of the term proportional to
$\Lambda^2_{\rm QCD}/\Lambda^2_{\rm U}$ differs the expression for the
gluon condensate obtained within the dilute--instanton gas
approximation from the gluon condensate calculated on lattice. The
appearance of this term is related to power corrections [13,21], whereas
the expression Eq.(\ref{label2.10}) corresponds to the leading order
contribution in power expansion [13]. The gluon condensate given by
Eqs.(\ref{label2.10}) and (\ref{label2.11}) we would compare with the
condensate of the dual--vector field $C_{\mu}$ which we calculate
below in the MNJL model.

\section{Dual--vector field condensate in the MNJL model}
\setcounter{equation}{0}

In the MNJL model we defined the condensate of the dual--vector field $C_{\mu}$
by analogy with the magnetic monopole condensate [2--4]
\begin{eqnarray}\label{label3.1}
\hspace{-0.5in}&&\Big\langle \frac{g^2}{4\pi^2}\,C_{\mu\nu}(x)\,
C^{\mu\nu}(x)\Big\rangle =
\frac{1}{Z}\int\!\!\!\int\!\!\!\int\!\!\!\int
D\chi\,D\bar{\chi}\,DC^{\mu}\,D\sigma\,\Bigg[\frac{g^2}{4\pi^2}\,
C_{\mu\nu}(x)\,C^{\mu\nu}(x)\Bigg]\,\nonumber\\
\hspace{-0.5in}&&\exp i\int
d^4z\,\Big[\frac{1}{2}\,C_{\nu}(z)\,(\Box + M^2_C)\,C^{\nu}(z) + C^{\nu}(z)\,\partial^{\mu}{^*{\cal E}_{\mu\nu}(z)}\nonumber\\
\hspace{-0.5in}&& +  \frac{1}{2}\,\sigma(z)\,(\Box + M^2_{\sigma})\,\sigma(z) + {\cal L}_{\rm int}[\sigma(z)]\nonumber\\
\hspace{-0.5in}&& -
g\,\bar{\chi}_M(z)\gamma_{\nu}\chi_M(z)\,C^{\nu}(z) -
\kappa\,\bar{\chi}_M(z)\chi_M(z)\,\sigma(z) +
\bar{\chi}_M(z)(i\,\gamma_{\nu}\partial^{\nu} - M)\,\chi_M(z)\Big],
\end{eqnarray}
where $Z$ is the normalization constant
\begin{eqnarray}\label{label3.2}
\hspace{-0.5in}&&Z =\int\!\!\!\int\!\!\!\int\!\!\!\int
D\chi\,D\bar{\chi}\,DC^{\mu}\,D\sigma\,\exp i\int
d^4z\,\Big[\frac{1}{2}\,C_{\nu}(z)\,(\Box +
M^2_C)\,C^{\nu}(z)\nonumber\\
\hspace{-0.5in}&& + \,C^{\nu}(z)\,\partial^{\mu}{^*{\cal E}_{\mu\nu}(z)} +
\frac{1}{2}\,\sigma(z)\,(\Box +
M^2_{\sigma})\,\sigma(z) + {\cal L}_{\rm int}[\sigma(z)]\nonumber\\
\hspace{-0.5in}&& -
g\,\bar{\chi}_M(z)\gamma_{\nu}\chi_M(z)\,C^{\nu}(z) -
\kappa\,\bar{\chi}_M(z)\chi_M(z)\,\sigma(z) +
\bar{\chi}_M(z)(i\,\gamma_{\nu}\partial^{\nu} - M)\,\chi_M(z)\Big].
\end{eqnarray}
Recall that we are working in the London limit, $M_{\sigma} = 2\,M \gg
M_C$. One can show that in this limit the contribution of the $\sigma$
field exchanges to the dual--vector field condensate are insignificant
and can be neglected with respect to the contributions of the
dual--vector field exchanges. As a result the integral over the
$\sigma$ field can be absorbed by the normalization constant $Z$
\begin{eqnarray}\label{label3.3}
&&\Big\langle \frac{g^2}{4\pi^2}\,C_{\mu\nu}(x)\,
C^{\mu\nu}(x)\Big\rangle = \frac{1}{Z}\int\!\!\!\int\!\!\!\int
D\chi\,D\bar{\chi}\,D C^{\mu}\,\Bigg[\frac{g^2}{4\pi^2}\,
C_{\mu\nu}(x)\,C^{\mu\nu}(x)\Bigg]\,\nonumber\\ &&\exp i\int
d^4z\,\Big[\frac{1}{2}\,C_{\nu}(z)\,(\Box + M^2_C)\,C^{\nu}(z)+
\,C^{\nu}(z)\,\partial^{\mu}{^*{\cal E}_{\mu\nu}(z)} -
g\,\bar{\chi}_M(z)\gamma_{\nu}\chi_M(z)\,C^{\nu}(z)\nonumber\\
\hspace{-0.5in}&& + \bar{\chi}_M(z)(i\,\gamma_{\nu}\partial^{\nu} - M)\,\chi_M(z)\Big].
\end{eqnarray}
The integration over the  dual--vector field $C_{\mu}$ we perform around
the Abrikosov flux line. For this aim we represent the dual--vector
field $C_{\mu}$ as follows [3,4,9,10]
\begin{eqnarray}\label{label3.4}
C_{\nu}(x) = C_{\nu}[{\cal E}(x)] + c_{\nu}(x).
\end{eqnarray}
The field $C_{\mu}[{\cal E}(x)]$ is the Abrikosov flux line
induced by a dual Dirac string obeying the equation [1--4,8--10]
\begin{eqnarray}\label{label3.5}
(\Box + M^2_C)\,C_{\nu}[{\cal E}(x)] = - \partial^{\mu}{^*{\cal
E}_{\mu\nu}(x)}
\end{eqnarray}
possessing the solution
\begin{eqnarray}\label{label3.6}
C_{\mu}[{\cal E}(x)] = - \int
d^4x'\,\Delta(x-x', M_C)\,\partial^{\mu}{^*{\cal E}_{\mu\nu}(x')},
\end{eqnarray}
where $\Delta(x-x', M_C)$ is the Green function
\begin{eqnarray}\label{label3.7}
\Delta(x-x', M_C) = \int \frac{d^4 k}{(2\pi)^4}\,\frac{\displaystyle
e^{\textstyle -i\,k\cdot(x-x') }}{M^2_C - k^2 - i\,0}.
\end{eqnarray}
The field $c_{\mu}(x)$ stands for the quantum fluctuations of the
dual--vector field around the Abrikosov flux line [2--4,9--10].
Denoting $\bar{F}_{\mu\nu}[{\cal E}(x)] = C_{\mu\nu}[{\cal E}(x)]$
and subtracting a trivial contribution proportional to
$\bar{F}_{\mu\nu}[{\cal E}(x)]\bar{F}^{\mu\nu}[{\cal E}(x)]$ we obtain
\begin{eqnarray}\label{label3.8}
\hspace{-0.3in}&&\delta \Big\langle \frac{g^2}{4\pi^2}\,
C_{\mu\nu}(x)\,C^{\mu\nu}(x)\Big\rangle = \Big\langle
\frac{g^2}{4\pi^2}\,C_{\mu\nu}(x)\,C^{\mu\nu}(x)\Big\rangle -
\frac{g^2}{4\pi^2}\,\bar{F}_{\mu\nu}[{\cal
E}(x)]\bar{F}^{\mu\nu}[{\cal
E}(x)]=\frac{1}{Z}\,\frac{g^2}{2\pi^2}\,\nonumber\\
\hspace{-0.3in}&&\times\int\!\!\!\int\!\!\!\int
D\chi\,D\bar{\chi}\,Dc^{\mu}\,\{\partial^{\mu}c^{\nu}(x)\,
(\partial_{\mu}c_{\nu}(x)
- \partial_{\nu}c_{\mu}(x)) + \bar{F}^{\mu\nu}[{\cal
E}(x)](\partial_{\mu}c_{\nu}(x) -
\partial_{\nu}c_{\mu}(x))\}\,\nonumber\\
\hspace{-0.3in}&&\exp i\int d^4z\,\Big[\frac{1}{2}\,c_{\nu}(z)\,(\Box
+ M^2_C)\,c^{\nu}(z) -
g\,\bar{\chi}_M(z)\gamma_{\nu}\chi_M(z)\,c^{\nu}(z) \nonumber\\
\hspace{-0.3in}&& -
g\,\bar{\chi}_M(z)\gamma_{\nu}\chi_M(z)\,C^{\nu}[{\cal E}(z)] +
\bar{\chi}_M(z)(i\,\gamma^{\nu}\partial_{\nu} - M)\,\chi_M(z)\Big].
\end{eqnarray}
For the integration over the $c_{\mu}$--field we suggest consider the
auxiliary path integral
\begin{eqnarray}\label{label3.9}
\hspace{-0.3in}&&{\cal J}(\bar{\chi}_M,\chi_M, C^{\nu}[{\cal E}])
=\nonumber\\
\hspace{-0.3in}&& = \int
Dc^{\mu}\,\{\partial_{\mu}c_{\nu}(x)\,(\partial_{\mu}c_{\nu}(x) -
\partial_{\nu}c_{\mu}(x)) + \bar{F}^{\mu\nu}[{\cal
E}(x)](\partial_{\mu}c_{\nu}(x) -
\partial_{\nu}c_{\mu}(x))\}\,\nonumber\\
\hspace{-0.3in}&&\exp i\int d^4z\,\Big[\frac{1}{2}\,c_{\nu}(z)\,(\Box
+ M^2_C)\,c^{\nu}(z) - c^{\nu}(z)\,J_{\nu}(z)\Big],
\end{eqnarray}
where $J_{\nu}(z)$ is a conserving current,
$\partial^{\nu}J_{\nu}(z) = 0$, defined by
\begin{eqnarray}\label{label3.10}
J_{\nu}(z) = g\,\bar{\chi}_M(z)\gamma_{\nu}\chi_M(z) + j_{\nu}(z)
\end{eqnarray}
and $j_{\nu}(z)$ is an external source which should be put zero
finally. 

The r.h.s. of Eq.(\ref{label3.9}) can be represented in the form of
functional derivatives with respect to $j_{\nu}(x)$ 
\begin{eqnarray}\label{label3.11}
\hspace{-0.3in}&&{\cal J}(\bar{\chi}_M,\chi_M, C^{\nu}[{\cal E}])
=\nonumber\\
\hspace{-0.3in}&&\Bigg\{-\frac{\partial}{\partial
x_{\mu}}\frac{\delta }{\delta
j_{\nu}(x)}\Bigg(\frac{\partial}{\partial x^{\mu}}\frac{\delta
}{\delta j^{\nu}(x)} - \frac{\partial}{\partial x^{\nu}}\frac{\delta
}{\delta j^{\mu}(x)}\Bigg) + i \bar{F}^{\mu\nu}[{\cal
E}(x)] \Bigg(\frac{\partial}{\partial x^{\mu}}\frac{\delta }{\delta
j^{\nu}(x)} - \frac{\partial}{\partial x^{\nu}}\frac{\delta }{\delta
j^{\mu}(x)}\Bigg)\Bigg\}\nonumber\\
\hspace{-0.3in}&&\times \int Dc^{\mu}\,\exp i\int
d^4z\,\Big[\frac{1}{2} c_{\nu}(z)\,(\Box + M^2_C) c^{\nu}(z) -
c^{\nu}(z) J_{\nu}(z)\Big]\Bigg|_{j_{\nu} = 0}.
\end{eqnarray}
At $J_{\nu}(z)=0$ the path integral over $c^{\mu}$ is normalized to
unity. The integral over $c^{\mu}$ is a Gaussian and the result of the
integration reads
\begin{eqnarray}\label{label3.12}
\hspace{-0.3in}&&\int Dc^{\mu}\,\exp i\int
d^4z\,\Big[\frac{1}{2}\,c_{\nu}(z)\,(\Box + M^2_C)\,c^{\nu}(z) -
c^{\nu}(z)\,J_{\nu}(z)\Big] = \nonumber\\
\hspace{-0.3in}&&= \exp\Big[ -i\,\frac{1}{2}\int\!\!\!\int
d^4z\,d^4z'\,J_{\nu}(z)\,\Delta(z - z', M_C)\,J^{\nu}(z')\Big],
\end{eqnarray}
where the Green function $\Delta(z - z', M_C)$ is given by
Eq.(\ref{label3.7}). 

The functional derivatives with respect to $j_{\nu}(x)$ are equal to
\begin{eqnarray}\label{label3.13}
\hspace{-0.5in}&&\frac{\partial }{\partial x^{\mu}}\frac{\delta }{\delta
j^{\nu}(x)}\exp\Big[ -i\,\frac{1}{2}\int\!\!\!\int
d^4z\,d^4z'\,J_{\nu}(z)\,\Delta(z - z', M_C)\,J^{\nu}(z')\Big]  =\nonumber\\
\hspace{-0.5in}&&= -\,i\int d^4z\,\frac{\partial }{\partial
x^{\mu}}\Delta(x - z, M_C)\,J_{\nu}(z) \nonumber\\
\hspace{-0.5in}&&\times\,\exp\Big[ -i\,\frac{1}{2}\int\!\!\!\int
d^4z\,d^4z'\,J_{\nu}(z)\,\Delta(z - z', M_C)\,J^{\nu}(z')\Big],\nonumber\\
\hspace{-0.5in}&&\frac{\partial }{\partial x^{\nu}}\frac{\delta }{\delta
j^{\mu}(x)}\exp\Big[ -i\,\frac{1}{2}\int\!\!\!\int
d^4z\,d^4z'\,J_{\nu}(z)\,\Delta(z - z', M_C)\,J^{\nu}(z')\Big]  =\nonumber\\
\hspace{-0.5in}&&= -\,i\int d^4z\,\frac{\partial }{\partial
x^{\nu}}\Delta(x - z, M_C)\,J_{\mu}(z) \nonumber\\
\hspace{-0.5in}&&\times\,\exp\Big[ -i\,\frac{1}{2}\int\!\!\!\int
d^4z\,d^4z'\,J_{\nu}(z)\,\Delta(z - z', M_C)\,J^{\nu}(z')\Big],\nonumber\\
\hspace{-0.5in}&&\frac{\partial }{\partial x_{\mu}}\frac{\delta }{\delta
j_{\nu}(x)}\frac{\partial }{\partial x^{\mu}}\frac{\delta }{\delta
j^{\nu}(x)}\exp\Big[ -i\,\frac{1}{2}\int\!\!\!\int
d^4z\,d^4z'\,J_{\nu}(z)\,\Delta(z - z', M_C)\,J^{\nu}(z')\Big]  =\nonumber\\
\hspace{-0.5in}&&= \Bigg[-\int
d^4z\,\Box_x\Delta(x - z, M_C)\,J_{\nu}(z)\int
d^4z'\,\Delta(x - z', M_C)\,J^{\nu}(z')\nonumber\\
\hspace{-0.5in}&& - \int d^4z\,\frac{\partial }{\partial
x_{\mu}}\Delta(x - z, M_C)\,J_{\nu}(z) \int d^4z'\,\frac{\partial
}{\partial x^{\mu}}\Delta(x - z', M_C)\,J^{\nu}(z')\Bigg] \nonumber\\
\hspace{-0.5in}&&\times\,\exp\Big[ -i\,\frac{1}{2}\int\!\!\!\int
d^4z\,d^4z'\,J_{\nu}(z)\,\Delta(z - z', M_C)\,J^{\nu}(z')\Big],\nonumber\\
\hspace{-0.5in}&&\frac{\partial }{\partial x_{\mu}}\frac{\delta }{\delta
j_{\nu}(x)}\frac{\partial }{\partial x^{\nu}}\frac{\delta }{\delta
j^{\mu}(x)}\exp\Big[ -i\,\frac{1}{2}\int\!\!\!\int
d^4z\,d^4z'\,J_{\nu}(z)\,\Delta(z - z', M_C)\,J^{\nu}(z')\Big]  =\nonumber\\
\hspace{-0.5in}&&= \Bigg[-\int d^4z\,\frac{\partial }{\partial
x^{\nu}}\Delta(x - z, M_C)\,J_{\mu}(z)\int d^4z'\,\frac{\partial
}{\partial x_{\mu}}\Delta(x - z', M_C)\,J^{\nu}(z')\Bigg] \nonumber\\
\hspace{-0.5in}&&\times\,\exp\Big[ -i\,\frac{1}{2}\int\!\!\!\int
d^4z\,d^4z'\,J_{\nu}(z)\,\Delta(z - z', M_C)\,J^{\nu}(z')\Big],
\end{eqnarray}
where we have taken into account that $\partial^{\mu}J_{\mu}(z) = 0$.

The functional ${\cal J}(\bar{\chi}_M,\chi_M, C^{\nu}[{\cal E}])$ is
then defined by
\begin{eqnarray}\label{label3.14}
\hspace{-0.3in}&&{\cal J}(\bar{\chi}_M,\chi_M, C^{\nu}[{\cal E}])=\nonumber\\
\hspace{-0.3in}&&=\Bigg\{g^2\int d^4z\,\Box_x\Delta(x - z,
M_C)[\bar{\chi}_M(z)\gamma_{\nu}\chi_M(z)]\int d^4z'\,\Delta(x - z',
M_C)[\bar{\chi}_M(z')\gamma^{\nu}\chi_M(z')]\nonumber\\ 
\hspace{-0.3in}&&+ g^2\int
d^4z\,\frac{\partial }{\partial x_{\mu}}\Delta(x - z,
M_C)\,[\bar{\chi}_M(z)\gamma_{\nu}\chi_M(z)] \int
d^4z'\,\frac{\partial }{\partial x^{\mu}}\Delta(x - z',
M_C)\,[\bar{\chi}_M(z')\gamma^{\nu}\chi_M(z')]\nonumber\\ 
\hspace{-0.3in}&&-\,g^2\int
d^4z\,\frac{\partial }{\partial x^{\nu}}\Delta(x - z,
M_C)\,[\bar{\chi}_M(z)\gamma_{\mu}\chi_M(z)] \int
d^4z'\,\frac{\partial }{\partial x_{\mu}}\Delta(x - z',
M_C)\,[\bar{\chi}_M(z')\gamma^{\nu}\chi_M(z')]\nonumber\\
\hspace{-0.3in}&&+¸\,2\,
g\,\bar{F}^{\mu\nu}[{\cal E}(x)]\int d^4z\,\frac{\partial
}{\partial x^{\mu}}\Delta(x - z,
M_C)\,[\bar{\chi}_M(z)\gamma_{\nu}\chi_M(z)]\Bigg\}\nonumber\\
\hspace{-0.3in}&&\times\,\exp\Big\{ -i\,g^2\,\frac{1}{2}\int\!\!\!\int
d^4z\,d^4z'\,[\bar{\chi}_M(z)\gamma_{\nu}\chi_M(z)]\Delta(z - z', M_C)\,
[\bar{\chi}_M(z')\gamma^{\nu}\chi_M(z')]\Big\}.
\end{eqnarray}
Substituting the functional ${\cal J}(\bar{\chi}_M,\chi_M,
C^{\nu}[{\cal E}])$ in the integrand of the r.h.s. of
Eq.(\ref{label3.8}) we express the dual--vector field condensate as a
path integral over the massive magnetic monopole fields.
\begin{eqnarray}\label{label3.15}
\hspace{-0.3in}&&\delta \Big\langle \frac{g^2}{4\pi^2}\,
C_{\mu\nu}(x)\,C^{\mu\nu}(x)\Big\rangle =\frac{1}{Z}\,\frac{g^2}{2\pi^2}\,\int\!\!\!\int\!\!\!\int
D\chi\,D\bar{\chi}\nonumber\\
\hspace{-0.3in}&&\Bigg\{g^2\int d^4z\,\Box_x\Delta(x - z,
M_C)[\bar{\chi}_M(z)\gamma_{\nu}\chi_M(z)]\int d^4z'\,\Delta(x - z',
M_C)[\bar{\chi}_M(z')\gamma^{\nu}\chi_M(z')]\nonumber\\ 
\hspace{-0.3in}&&+ g^2\int
d^4z\,\frac{\partial }{\partial x_{\mu}}\Delta(x - z,
M_C)\,[\bar{\chi}_M(z)\gamma_{\nu}\chi_M(z)] \int
d^4z'\,\frac{\partial }{\partial x^{\mu}}\Delta(x - z',
M_C)\,[\bar{\chi}_M(z')\gamma^{\nu}\chi_M(z')]\nonumber\\ 
\hspace{-0.3in}&&-\,g^2\int
d^4z\,\frac{\partial }{\partial x^{\nu}}\Delta(x - z,
M_C)\,[\bar{\chi}_M(z)\gamma_{\mu}\chi_M(z)] \int
d^4z'\,\frac{\partial }{\partial x_{\mu}}\Delta(x - z',
M_C)\,[\bar{\chi}_M(z')\gamma^{\nu}\chi_M(z')]\nonumber\\
\hspace{-0.3in}&&+¸\,2\,
g\,\bar{F}^{\mu\nu}[{\cal E}(x)]\int d^4z\,\frac{\partial
}{\partial x^{\mu}}\Delta(x - z,
M_C)\,[\bar{\chi}_M(z)\gamma_{\nu}\chi_M(z)]\Bigg\}\nonumber\\
\hspace{-0.3in}&&\times\,\exp i\int
d^4z\,\Big\{\bar{\chi}_M(z)(i\,\gamma^{\nu}\partial_{\nu} -
M)\,\chi_M(z) - g\,\bar{\chi}_M(z)\gamma_{\nu}\chi_M(z)\,C^{\nu}[{\cal
E}(z)]\nonumber\\ 
\hspace{-0.3in}&&-\,i\,g^2\,\frac{1}{2}\int\!\!\!\int
d^4z\,d^4z'\,[\bar{\chi}_M(z)\gamma_{\nu}\chi_M(z)]\Delta(z - z', M_C)\,
[\bar{\chi}_M(z')\gamma^{\nu}\chi_M(z')]\Big\}.
\end{eqnarray}
As has been shown in Refs.[3,4] due to the London limit, $M\gg M_C$,
the non--local four--monopole interaction in the exponential of the
r.h.s. of Eq.(\ref{label3.15}) can be approximated by a local
four--monopole interactions. This reduces the r.h.s. of
Eq.(\ref{label3.15}) to the form [3,4]
\begin{eqnarray}\label{label3.16}
\hspace{-0.3in}&&\delta \Big\langle \frac{g^2}{4\pi^2}\,
C_{\mu\nu}(x)\,C^{\mu\nu}(x)\Big\rangle =
\frac{1}{Z}\,\frac{g^2}{2\pi^2}\,\int\!\!\!\int
D\chi\,D\bar{\chi}\nonumber\\
\hspace{-0.3in}&&\Bigg\{g^2\int d^4z\,\Box_x\Delta(x - z,
M_C)[\bar{\chi}_M(z)\gamma_{\nu}\chi_M(z)]\int d^4z'\,\Delta(x - z',
M_C)[\bar{\chi}_M(z')\gamma^{\nu}\chi_M(z')]\nonumber\\ 
\hspace{-0.3in}&&+ g^2\int
d^4z\,\frac{\partial }{\partial x_{\mu}}\Delta(x - z,
M_C)\,[\bar{\chi}_M(z)\gamma_{\nu}\chi_M(z)] \int
d^4z'\,\frac{\partial }{\partial x^{\mu}}\Delta(x - z',
M_C)\,[\bar{\chi}_M(z')\gamma^{\nu}\chi_M(z')]\nonumber\\ 
\hspace{-0.3in}&&-\,g^2\int
d^4z\,\frac{\partial }{\partial x^{\nu}}\Delta(x - z,
M_C)\,[\bar{\chi}_M(z)\gamma_{\mu}\chi_M(z)] \int
d^4z'\,\frac{\partial }{\partial x_{\mu}}\Delta(x - z',
M_C)\,[\bar{\chi}_M(z')\gamma^{\nu}\chi_M(z')]\nonumber\\
\hspace{-0.3in}&&+¸\,2\,
g\,\bar{F}^{\mu\nu}[{\cal E}(x)]\int d^4z\,\frac{\partial
}{\partial x^{\mu}}\Delta(x - z,
M_C)\,[\bar{\chi}_M(z)\gamma_{\nu}\chi_M(z)]\Bigg\}\nonumber\\
\hspace{-0.3in}&&\times\,\exp i\int
d^4z\,\Big\{\bar{\chi}_M(z)(i\,\gamma^{\nu}\partial_{\nu} -
M)\,\chi_M(z) - g\,\bar{\chi}_M(z)\gamma_{\nu}\chi_M(z)\,C^{\nu}[{\cal
E}(z)]\nonumber\\ 
\hspace{-0.3in}&&-
\frac{g^2}{2M^2_C}[\bar{\chi}_M(z)\gamma_{\nu}\chi_M(z)]
[\bar{\chi}_M(z)\gamma^{\nu}\chi_M(z)]\Big\}.
\end{eqnarray}
For the integration over the massive magnetic monopole fields it is
convenient to decompose the r.h.s. of Eq.(\ref{label3.16})
conventionally into two parts
\begin{eqnarray}\label{label3.17}
\delta \Big\langle \frac{g^2}{4\pi^2}\,C_{\mu\nu}(x)\,
C^{\mu\nu}(x)\Big\rangle &=& \Big\langle \frac{g^2}{4\pi^2}\,
C_{\mu\nu}(x)\, C^{\mu\nu}(x)\Big\rangle_{\rm non-string}\nonumber\\
&+& 
\Big\langle \frac{g^2}{4\pi^2}\, C_{\mu\nu}(x)\,
C^{\mu\nu}(x)\Big\rangle_{\rm string},
\end{eqnarray}
where the terms are given by
\begin{eqnarray}\label{label3.18}
\hspace{-0.3in}&&\Big\langle \frac{g^2}{4\pi^2}\,
C_{\mu\nu}(x)\,C^{\mu\nu}(x)\Big\rangle_{\rm non-string} = 
\frac{1}{Z}\,\frac{g^2}{2\pi^2}\,\int\!\!\!\int
D\chi\,D\bar{\chi}\nonumber\\
\hspace{-0.3in}&&\Bigg\{g^2\int d^4z\,\Box_x\Delta(x - z,
M_C)[\bar{\chi}_M(z)\gamma_{\nu}\chi_M(z)]\int d^4z'\,\Delta(x - z',
M_C)[\bar{\chi}_M(z')\gamma^{\nu}\chi_M(z')]\nonumber\\ 
\hspace{-0.3in}&&+ g^2\int
d^4z\,\frac{\partial }{\partial x_{\mu}}\Delta(x - z,
M_C)\,[\bar{\chi}_M(z)\gamma_{\nu}\chi_M(z)] \int
d^4z'\,\frac{\partial }{\partial x^{\mu}}\Delta(x - z',
M_C)\,[\bar{\chi}_M(z')\gamma^{\nu}\chi_M(z')]\nonumber\\ 
\hspace{-0.3in}&&-\,g^2\int
d^4z\,\frac{\partial }{\partial x^{\nu}}\Delta(x - z,
M_C)\,[\bar{\chi}_M(z)\gamma_{\mu}\chi_M(z)] \int
d^4z'\,\frac{\partial }{\partial x_{\mu}}\Delta(x - z',
M_C)\,[\bar{\chi}_M(z')\gamma^{\nu}\chi_M(z')]\nonumber\\
\hspace{-0.3in}&&\times\,\exp i\int
d^4z\,\Big\{\bar{\chi}_M(z)(i\,\gamma^{\nu}\partial_{\nu} -
M)\,\chi_M(z) - g\,\bar{\chi}_M(z)\gamma_{\nu}\chi_M(z)\,C^{\nu}[{\cal
E}(z)]\nonumber\\ 
\hspace{-0.3in}&&-
\frac{g^2}{2M^2_C}[\bar{\chi}_M(z)\gamma_{\nu}\chi_M(z)]
[\bar{\chi}_M(z)\gamma^{\nu}\chi_M(z)]\Big\}
\end{eqnarray}
and 
\begin{eqnarray}\label{label3.19}
\hspace{-0.3in}&&\Big\langle \frac{g^2}{4\pi^2}\,
C_{\mu\nu}(x)\,C^{\mu\nu}(x)\Big\rangle_{\rm string} = 
\frac{1}{Z}\,\frac{g^2}{\pi^2}\,\bar{F}^{\mu\nu}[{\cal
E}(x)] \int\!\!\!\int D\chi\,D\bar{\chi}\nonumber\\
\hspace{-0.3in}&&\times \int
d^4z\,\frac{\partial }{\partial x^{\mu}}\Delta(x - z,
M_C)\,g\,[\bar{\chi}_M(z)\gamma_{\nu}\chi_M(z)]\nonumber\\
\hspace{-0.3in}&&\times\,\exp i\int
d^4z\,\Big\{\bar{\chi}_M(z)(i\,\gamma^{\nu}\partial_{\nu} -
M)\,\chi_M(z)- g\,\bar{\chi}_M(z)\gamma_{\nu}\chi_M(z)\,C^{\nu}[{\cal
E}(z)] \nonumber\\
\hspace{-0.3in}&& - \frac{g^2}{2M^2_C}[\bar{\chi}_M(z)\gamma_{\nu}\chi_M(z)]
[\bar{\chi}_M(z)\gamma^{\nu}\chi_M(z)]\Big\}.
\end{eqnarray}
As we would show below the non--string part of the dual--vector field
condensate does not depend on a dual Dirac string, whereas the string
part does.

The integration over massive monopole fields is analogous
to that carried out in Refs.[3--4], where the r.h.s. of
Eq.(\ref{label3.19}) has been approximated by massive monopole--loop
diagrams by keeping only leading divergent contributions. 

In terms of the massive magnetic monopole--loop diagrams the
non--string part of the dual--vector field condensate is determined by
\begin{eqnarray}\label{label3.20}
\hspace{-0.3in}&&\Big\langle \frac{g^2}{4\pi^2}\,C_{\mu\nu}(x)\,
C^{\mu\nu}(x)\Big\rangle_{\rm non-string}
=\Bigg(\frac{g^2}{16\pi^2}\Bigg)^2\int\frac{d^4q}{2\pi^4i}\,
\frac{q^{\mu}q^{\nu}}{(M^2_C
- q^2 - i\,0)^2}\nonumber\\
\hspace{-0.3in}&&\Bigg[\int\frac{d^4k}{\pi^2i}\, {\rm
tr}\,\Bigg\{\gamma_{\mu}\frac{1}{M - \hat{k} + g\,\hat{C}[{\cal
E}(z)]}\gamma_{\nu}\frac{1}{M - \hat{k} + g\,\hat{C}[{\cal
E}(z)]}\Bigg\} + \sum^{\infty}_{n =
1}\Bigg[\frac{g^2}{M^2_C}\Bigg]^n\Bigg(\frac{1}{16\pi^2}\Bigg)^n
\nonumber\\
\hspace{-0.3in}&& \nonumber\\
\hspace{-0.3in}&&\times \int\frac{d^4k_1}{\pi^2i}\,
{\rm tr}\,\Bigg\{\gamma_{\mu}\frac{1}{M - \hat{k}_1 + g\,\hat{C}[{\cal
E}(z)]}\gamma_{\alpha_1}\frac{1}{M - \hat{k}_1 + g\,\hat{C}[{\cal
E}(z)]}\Bigg\}\nonumber\\
\hspace{-0.3in}&&\times\int\frac{d^4k_2}{\pi^2i}\,
{\rm tr}\,\Bigg\{\gamma_{\alpha_1}\frac{1}{M - \hat{k}_2 + g\,\hat{C}[{\cal
E}(z)]}\gamma_{\alpha_2}\frac{1}{M - \hat{k}_2 + g\,\hat{C}[{\cal
E}(z)]}\Bigg\}\ldots\nonumber\\
\hspace{-0.3in}&&\times \int\frac{d^4k_n}{\pi^2i}\, {\rm
tr}\,\Bigg\{\gamma_{\alpha_{n-1}}\frac{1}{M - \hat{k}_n +
g\,\hat{C}[{\cal E}(z)]}\gamma_{\alpha_n}\frac{1}{M - \hat{k}_n +
g\,\hat{C}[{\cal E}(z)]}\Bigg\}\nonumber\\
\hspace{-0.3in}&&\times
\int\frac{d^4k_{n+1}}{\pi^2i}\, {\rm
tr}\,\Bigg\{\gamma_{\alpha_n}\frac{1}{M - \hat{k}_{n+1} + g\,\hat{C}[{\cal
E}(z)]}\gamma_{\nu}\frac{1}{M - \hat{k}_{n+1} + g\,\hat{C}[{\cal
E}(z)]}\Bigg\}\Bigg].
\end{eqnarray}
Here we have taken into account that the contributions of the first
two terms in the non--string part cancel each other after integration
over virtual momenta.

Integrating over $k$ and $k_i\,(i = 1,\ldots,n + 1)$ we obtain [3,4]
\begin{eqnarray}\label{label3.21}
\hspace{-0.3in}&&\Big\langle \frac{g^2}{4\pi^2}\,C_{\mu\nu}(x)\,
C^{\mu\nu}(x)\Big\rangle_{\rm non-string}
=\Bigg(\frac{g^2}{16\pi^2}\Bigg)^2\int\frac{d^4q}{2\pi^4i}\,
\frac{q^{\mu}q^{\nu}}{(M^2_C
- q^2 - i\,0)^2}\nonumber\\
\hspace{-0.3in}&&2\,g_{\mu\nu}\,[J_1(M) + M^2J_2(M)]\sum^{\infty}_{n =
1}\Bigg(\frac{1}{M^2_C}\frac{g^2}{8\pi^2}[J_1(M) +
M^2J_2(M)]\Bigg)^n=\nonumber\\
\hspace{-0.3in}&&=\frac{g^2}{32\pi^4}\,[J_1(M_C) -
M^2_CJ_2(M_C)]\frac{\displaystyle
\frac{g^2}{8\pi^2}[J_1(M) +
M^2J_2(M)]}{\displaystyle 1 -
\frac{1}{M^2_C}\,\frac{g^2}{8\pi^2}\,[J_1(M) + M^2J_2(M)]},
\end{eqnarray}
where $J_1(M_C)$ and $J_2(M_C)$ are the quadratically and
logarithmically divergent integrals defined by Eq.(\ref{label1.15})
with $M$ replaced by $M_C$.  Since the cut--off $\Lambda$ is much
greater than $M_C$, we can replace $[J_1(M_C) - M^2_CJ_2(M_C)]$ by
$\Lambda^2$. The non--string part of the dual--vector field condensate
is then given by
\begin{eqnarray}\label{label3.22}
\hspace{-0.3in}\frac{1}{\Lambda^4}\Big\langle \frac{g^2}{4\pi^2}\,
C_{\mu\nu}(x)\,C^{\mu\nu}(x)\Big\rangle_{\rm non-string}
=\frac{1}{\Lambda^2}\,\frac{g^2}{32\pi^4}\,\frac{\displaystyle
\frac{g^2}{8\pi^2}[J_1(M) +
M^2J_2(M)]}{\displaystyle 1 -
\frac{1}{M^2_C}\,\frac{g^2}{8\pi^2}\,[J_1(M) + M^2J_2(M)]}.
\end{eqnarray}
We have represented the non--string part of the dual--vector field
condensate in the form convenient for the comparison with the lattice
calculations where we have set $\Lambda_{\rm U} = \Lambda$.

The string part of the dual--vector field condensate is determined by
the following expression in terms of the massive magnetic
monopole--loop diagrams [3,4]
\begin{eqnarray}\label{label3.23}
\hspace{-0.3in}&&\Big\langle \frac{g^2}{4\pi^2}\,
C_{\mu\nu}(x)\, C^{\mu\nu}(x)\Big\rangle_{\rm string} =
\frac{g^2}{\pi^2}\,\bar{F}^{\mu\nu}[{\cal E}(x)]\,\int d^4z\,\frac{\partial
}{\partial x^{\mu}}\Delta(x - z, M_C)\,\nonumber\\
\hspace{-0.3in}&&\Bigg[\Bigg(-\frac{g}{16\pi^2}\Bigg)\int\frac{d^4k}{\pi^2i}\,
{\rm tr}\,\Bigg\{\gamma_{\nu}\frac{1}{M - \hat{k} + g\,\hat{C}[{\cal
E}(z)]}\Bigg\} + \sum^{\infty}_{n =
1}\Bigg[\frac{g^2}{M^2_C}\Bigg]^n\Bigg(\frac{1}{16\pi^2}\Bigg)^n \nonumber\\
\hspace{-0.3in}&&\times \int\frac{d^4k_1}{\pi^2i}\,
{\rm tr}\,\Bigg\{\gamma_{\nu}\frac{1}{M - \hat{k}_1 + g\,\hat{C}[{\cal
E}(z)]}\gamma_{\alpha_1}\frac{1}{M - \hat{k}_1 + g\,\hat{C}[{\cal
E}(z)]}\Bigg\}\nonumber\\
\hspace{-0.3in}&&\times\int\frac{d^4k_2}{\pi^2i}\,
{\rm tr}\,\Bigg\{\gamma_{\alpha_1}\frac{1}{M - \hat{k}_2 + g\,\hat{C}[{\cal
E}(z)]}\gamma_{\alpha_2}\frac{1}{M - \hat{k}_2 + g\,\hat{C}[{\cal
E}(z)]}\Bigg\}\ldots\nonumber\\
\hspace{-0.3in}&&\times \int\frac{d^4k_{n-1}}{\pi^2i}\, {\rm
tr}\,\Bigg\{\gamma_{\alpha_{n-1}}\frac{1}{M - \hat{k}_{n-1} +
g\,\hat{C}[{\cal E}(z)]}\gamma_{\alpha_n}\frac{1}{M - \hat{k}_{n-1} +
g\,\hat{C}[{\cal E}(z)]}\Bigg\}\nonumber\\
\hspace{-0.3in}&&\times\Bigg(-\frac{g}{16\pi^2}\Bigg)
\int\frac{d^4k_n}{\pi^2i}\, {\rm
tr}\,\Bigg\{\gamma_{\alpha_n}\frac{1}{M - \hat{k}_n + g\,\hat{C}[{\cal
E}(z)]}\Bigg\}\Bigg].
\end{eqnarray}
Keeping only leading divergent contributions [2--4] we obtain
\begin{eqnarray}\label{label3.24}
\hspace{-0.3in}&&\Big\langle \frac{g^2}{4\pi^2}\,
C_{\mu\nu}(x)\, C^{\mu\nu}(x)\Big\rangle_{\rm string} =
\frac{g^2}{\pi^2}\,\bar{F}^{\mu\nu}[{\cal E}(x)]\,\int d^4z\,\frac{\partial
}{\partial x^{\mu}}\Delta(x - z, M_C)\,C_{\nu}[{\cal E}(z)]\nonumber\\
\hspace{-0.3in}&&\times\,\frac{\displaystyle
\frac{g^2}{8\pi^2}\,[J_1(M) + M^2J_2(M)]}{\displaystyle 1 -
\frac{1}{M^2_C}\,\frac{g^2}{8\pi^2}\,[J_1(M) + M^2J_2(M)]}.
\end{eqnarray}
Integrating by parts over $z$ we arrive at the expression
\begin{eqnarray}\label{label3.25}
\hspace{-0.3in}&&\Big\langle \frac{g^2}{4\pi^2}\,C_{\mu\nu}(x)\,
C^{\mu\nu}(x)\Big\rangle_{\rm string} = \nonumber\\
\hspace{-0.3in}&&=\frac{\displaystyle
\frac{g^2}{8\pi^2}\,[J_1(M) + M^2J_2(M)]}{\displaystyle 1 -
\frac{1}{M^2_C}\,\frac{g^2}{8\pi^2}\,[J_1(M) + M^2J_2(M)]}\int
d^4z\,\frac{g^2}{2\pi^2}\,\bar{F}^{\mu\nu}[{\cal E}(x)]\,\Delta(x - z,
M_C)\,\bar{F}_{\mu\nu}[{\cal E}(z)].
\end{eqnarray}
Collecting all pieces we obtain the condensate of the dual--vector
field 
\begin{eqnarray}\label{label3.26}
\hspace{-0.3in}&&\frac{1}{\Lambda^4}\Big\langle \frac{g^2}{4\pi^2}\,
C_{\mu\nu}(x)\, C^{\mu\nu}(x)\Big\rangle =\frac{1}{\Lambda^4}\,
\frac{g^2}{4\pi^2}\,\bar{F}^{\mu\nu}[{\cal
E}(x)]\,\bar{F}_{\mu\nu}[{\cal E}(x)] + \frac{\displaystyle
\frac{g^2}{8\pi^2}[J_1(M) + M^2J_2(M)]}{\displaystyle 1 -
\frac{1}{M^2_C}\,\frac{g^2}{8\pi^2}\,[J_1(M) + M^2J_2(M)]}\nonumber\\
\hspace{-0.3in}&&\times\Bigg(\frac{1}{\Lambda^2}\,\frac{g^2}{32\pi^4}\,
+ \,\frac{1}{\Lambda^4}\,\int
d^4z\,\frac{g^2}{2\pi^2}\,\bar{F}^{\mu\nu}[{\cal E}(x)]\,\Delta(x - z,
M_C)\,\bar{F}_{\mu\nu}[{\cal E}(z)]\Bigg),
\end{eqnarray}
where the first term comes from the tree--approximation defined by
the Abrikosov flux line, whereas the second one is fully caused by
quantum field fluctuations around the Abrikosov flux line. Below we
show that the term caused by quantum field fluctuations dominates in
the dual--vector field condensate.

By using Eqs.(\ref{label1.19})--(\ref{label1.17}) and the relation
$G_1 = G/4$ we can recast the r.h.s. of Eq.(\ref{label3.26}) into the
form
\begin{eqnarray}\label{label3.27}
\hspace{-0.3in}&&\frac{1}{\Lambda^4}\Big\langle \frac{g^2}{4\pi^2}\,
C_{\mu\nu}(x)\, C^{\mu\nu}(x)\Big\rangle =\frac{1}{\Lambda^4}\,
\frac{g^2}{4\pi^2}\,\bar{F}^{\mu\nu}[{\cal
E}(x)]\,\bar{F}_{\mu\nu}[{\cal E}(x)] + \frac{\displaystyle 1 -
\frac{g^2}{3}\,\frac{\langle \bar{\chi}\chi\rangle}{M^3}}{\displaystyle 1 +
\frac{2 g^2}{M^2_C}\,\frac{\langle \bar{\chi}\chi\rangle}{M}}\nonumber\\
\hspace{-0.3in}&&\times\Bigg(\frac{M^2}{\Lambda^2}\,\frac{3 g^2}{128\pi^4}\,
+ \,\frac{M^2}{\Lambda^4}\,\int
d^4z\,\frac{3 g^2}{8\pi^2}\,\bar{F}^{\mu\nu}[{\cal E}(x)]\,\Delta(x - z,
M_C)\,\bar{F}_{\mu\nu}[{\cal E}(z)]\Bigg).
\end{eqnarray}
The dependence of the dual--vector field condensate on the shape of a
dual Dirac string enters through the term $\bar{F}^{\mu\nu}[{\cal
E}(x)]\,\bar{F}_{\mu\nu}[{\cal E}(x)]$.

Since the confinement regime can be described by static infinitely
long dual Dirac strings, we would make an estimate of the numerical
value of the dual--vector field condensate in the approximation of the
infinitely long dual Dirac string strained along the $z$--axis. In
this approximation the electric field of a static dual Dirac string is given
by [8]
\begin{eqnarray}\label{label3.28}
\vec{\cal E}(\vec{r}\,) = \vec{e}_z\,Q\,\delta^{(2)}(\vec{r}\,),
\end{eqnarray}
where $\vec{r}$ is the radius--vector in the plane perpendicular to
the $z$--axis. The dual--vector potential possesses only the azimuthal
component and reads [8]
\begin{eqnarray}\label{label3.29}
C_{\alpha}(r) = -\,\frac{QM_C}{2\pi}\,K_1(M_Cr),
\end{eqnarray}
where $K_1(M_Cr)$ is the McDonald function. The dual electric field
induced by an infinitely long dual Dirac string amounts to
\begin{eqnarray}\label{label3.30}
\vec{E}(r) = {\rm
rot}\,\vec{C}(\vec{r}\,)= -\,\vec{e}_z\,\frac{QM^2_C}{2\pi}\,K_0(M_Cr).
\end{eqnarray}
Substituting the dual electric field Eq.(\ref{label3.30}) in the dual--vector
field condensate Eq.(\ref{label3.26}) we obtain
\begin{eqnarray}\label{label3.31}
\hspace{-0.3in}&&\frac{1}{\Lambda^4}\Big\langle \frac{g^2}{4\pi^2}\,
C_{\mu\nu}(x)\, C^{\mu\nu}(x)\Big\rangle =\frac{1}{\Lambda^4}\,
\frac{g^2}{2\pi^2}\,\frac{Q^2M^4_C}{4\pi^2}\,K^2_0(M_Cr) +
\frac{\displaystyle 1 - \frac{g^2}{3}\,\frac{\langle
\bar{\chi}\chi\rangle}{M^3}}{\displaystyle 1 + \frac{2
g^2}{M^2_C}\,\frac{\langle \bar{\chi}\chi\rangle}{M}}\nonumber\\
\hspace{-0.3in}&&\times\Bigg(\frac{M^2}{\Lambda^2}\,\frac{3
g^2}{128\pi^4}\, +
\,\frac{M^2M^2_C}{\Lambda^4}\,\frac{3
g^2}{8\pi^2}\,\frac{Q^2}{4\pi^2}\,M_cr\,K_0(M_Cr)\,K_1(M_Cr)\Bigg).
\end{eqnarray}
By using the Dirac quantization condition $g\,Q = 2\pi$ we bring up  the
r.h.s. of Eq.(\ref{label3.32}) to the form
\begin{eqnarray}\label{label3.32}
\hspace{-0.3in}&&\frac{1}{\Lambda^4}\Big\langle \frac{g^2}{4\pi^2}\,
C_{\mu\nu}(x)\, C^{\mu\nu}(x)\Big\rangle =\frac{M^4_C}{\Lambda^4}\,
\frac{1}{2\pi^2}\,K^2_0(M_Cr) +
\frac{\displaystyle 1 - \frac{g^2}{3}\,\frac{\langle
\bar{\chi}\chi\rangle}{M^3}}{\displaystyle 1 + \frac{2
g^2}{M^2_C}\,\frac{\langle \bar{\chi}\chi\rangle}{M}}\nonumber\\
\hspace{-0.3in}&&\times\Bigg(\frac{M^2}{\Lambda^2}\,\frac{3
g^2}{128\pi^4}\, +
\,\frac{M^2M^2_C}{\Lambda^4}\,\frac{3}{8\pi^2}\,M_cr\,K_0(M_Cr)
\,K_1(M_Cr)\Bigg).
\end{eqnarray}
At $r\to \infty$ the dual--vector field condensate is defined by only
the non--string part and reads
\begin{eqnarray}\label{label3.33}
\frac{1}{\Lambda^4}\Big\langle \frac{g^2}{4\pi^2}\, C_{\mu\nu}(x)\,
C^{\mu\nu}(x)\Big\rangle =\frac{M^2}{\Lambda^2}\,\frac{3\,g^2}{128\pi^4}\,\frac{\displaystyle 1 -
\frac{g^2}{3}\,\frac{\langle \bar{\chi}\chi\rangle}{M^3}}{\displaystyle 1 + 2\,g^2\,\frac{\langle \bar{\chi}\chi\rangle}{M^2_CM}}.
\end{eqnarray}
One can show that this value is positive.

At $r\to 0$ the dual--vector field condensate is infinite. However, as
has been noted in Refs.[2--4] in the MNJL model as well as in the dual
Higgs model with dual Dirac strings [8--10] the minimal transversal
distances are restricted by inequality $r\ge 1/M_{\sigma} =
1/2M$. Thus, at $r =1/2M$ we get the maximal value of the dual--vector
field condensate, since the contributions of the dual Dirac string
described by the terms proportional to $K^2_0(MCr)$ and
$r\,K_0(M_Cr)\,K_1(M_Cr)$ are positive.
\begin{eqnarray}\label{label3.34}
\hspace{-0.3in}&&\frac{1}{\Lambda^4}\Big\langle \frac{g^2}{4\pi^2}\,
C_{\mu\nu}(x)\, C^{\mu\nu}(x)\Big\rangle =\frac{M^4_C}{\Lambda^4}\,
\frac{1}{2\pi^2}\,{\ell n}^2\Bigg(\frac{M^2}{M^2_C}\Bigg) +
\frac{\displaystyle 1 - \frac{g^2}{3}\,\frac{\langle
\bar{\chi}\chi\rangle}{M^3}}{\displaystyle 1 + \frac{2
g^2}{M^2_C}\,\frac{\langle \bar{\chi}\chi\rangle}{M}}\nonumber\\
\hspace{-0.3in}&&\times\Bigg(\frac{M^2}{\Lambda^2}\,\frac{3
g^2}{128\pi^4}\, +
\,\frac{M^2M^2_C}{\Lambda^4}\,\frac{3}{8\pi^2}\,{\ell
n}\Bigg(\frac{M^2}{M^2_C}\Bigg)\Bigg).
\end{eqnarray}
We should emphasize that unlike the magnetic monopole condensate
the dual--vector field condensate does not vanish in the close
vicinity of a dual Dirac string [3,4].

\section{Conclusion}
\setcounter{equation}{0}

The evaluation of the dual--vector field condensate in the MNJL model
has confirmed our statement we have pointed out in Refs.[2--4,9,10]
that quantum fluctuations of the fields of monopole--(anti)monopole
collective excitations around the Abrikosov flux lines induced by dual
Dirac strings in the condensed phase play an important role for the
undestanding of the confinement mechanism. We have shown that these
fluctuations determine fully the dual--vector field condensate
independent on the shape of a dual Dirac string and non--vanishing at
large distances.

In the case of the neglect of quantum field fluctuations this part of
the dual--vector field condensate does not appear and the condensate
is defined by
\begin{eqnarray}\label{label4.1}
\frac{1}{\Lambda^4}\Big\langle \frac{g^2}{4\pi^2}\, C_{\mu\nu}(x)\,
C^{\mu\nu}(x)\Big\rangle = \frac{1}{\Lambda^4}\,\frac{g^2}{4\pi^2}\,\bar{F}^{\mu\nu}[{\cal
E}(x)]\,\bar{F}_{\mu\nu}[{\cal E}(x)] =
\frac{M^4_C}{\Lambda^4}\,\frac{1}{2\pi^2}\,K^2_0(M_Cr)
\end{eqnarray}
for the infinitely long static dual Dirac string strained along the
$z$--axis.

Unlike Eq.(\ref{label3.32}) the dual--vector field condensate given by
Eq.(\ref{label4.1}) vanishes at large distances in the plane
transversal to a dual Dirac string. This result should contradict to
the gluon condensate having a non--vanishing part at large distances.

When matching our expression for the dual--vector field condensate
given by Eq.(\ref{label3.32}) with that calculated on lattice we would
like to emphasize the absence of the perturbative term $W_{40}$ and
the presence of the term proportional to $M^2/\Lambda^2$ having a
non--perturbative nature. The absence of the perturbative contribution
to the dual--vector field condensate is rather clear. Indeed, the
evaluation of the dual--vector field condensate is carried out in the
non--perturbative (condensed) phase of the MNJL model with a
non--perturbative dual--superconducting vacuum filled with dual Dirac
strings. Thereby, the perturbative contributions cannot appear in
principle.

Then, non--perturbative contributions having the structure
$M^4_C/\Lambda^4$ and $M^2_CM^2/\Lambda^4$ are determined fully by the
dual Dirac string. They vanish at large distances. At short distances
restricted from below by inequality $r \ge 1/M_{\sigma} = 1/2M$ the
contributions of these terms are positive. Due to this
they increase the value of the non--string part of the dual--vector
field condensate in the vicinity of a dual Dirac string. This result
is opposite to that we have obtained for the magnetic monopole
condensate.  Unlike the dual--vector field condensate the magnetic
monopole condensate vanishes in the close vicinity of a dual Dirac
string [2--4].

Unfortunately, we could not derive with one--to--one correspondence
the contribution to the dual--vector field condensate analogous to
that caused by a dilute--instanton vacuum in the gluon condensate. In
the MNJL model according to the Dirac quantization this contribution
should be of order $(M^4_C/\Lambda^4)\,O(g^6)$ or
$(M^4_C/\Lambda^4)\,O(g^6)$. Such contributions can be found among
other terms in the part of the dual--vector field condensate depending
on the shape of a dual Dirac string. However, these terms vanish at
large distances and do not give a dominant contribution to the
dual--vector field condensate. Thus, we can conclude that the
non--perturbative dual--superconducting vacuum of the MNJL model
possesses partly the properties of a dilute--instanton gas vacuum but
this part of the wave function of the non--perturbative vacuum does
not play a dominant role.

\section{Acknowledgement}

We are grateful to Prof. M. I. Polikarpov for helpful discussions and
comments.

\end{document}